# Numerical solution to the Bloch equations: paramagnetic solutions under wideband continuous radio frequency irradiation in a pulsed magnetic field

**Wen-Jun Chen(陈文俊)[1], Hong Ma(马洪)[2], De Yu(余德)[1], Xiao-Hu Zeng(曾小虎)[2]**

[1] School of Physics, Huazhong University of Science and Technology, Wuhan 430074, China

[2] School of Electronic Information and Communications, Huazhong University of Science and Technology, Wuhan 430074, China

**Abstract** A novel nuclear magnetic resonance (NMR) experimental scheme, called wideband continuous wave NMR (WB-CW-NMR), is presented in this article. This experimental scheme has promising applications in pulsed magnetic fields, and can dramatically improve the utilization of the pulsed field. The feasibility of WB-CW-NMR scheme is verified by numerically solving modified Bloch equations. In the numerical simulation, the applied magnetic field is a pulsed magnetic field up to 80 T, and the wideband continuous radio frequency (RF) excitation is a band-limited (0.68–3.40 GHz) white noise. Furthermore, the influences of some experimental parameters, such as relaxation time, applied magnetic field strength and wideband continuous RF power, on the WB-CW-NMR signal are analyzed briefly. Finally, a multi-channel system framework for transmitting and receiving ultra wideband signals is proposed, and the basic requirements of this experimental system are discussed. Meanwhile, the amplitude of the NMR signal, the level of noise and RF interference in WB-CW-NMR experiments are estimated, and a preliminary adaptive cancellation plan is given for detecting WB-CW-NMR signal from large background interference.

# 1. Introduction

Among various measurements, nuclear magnetic resonance (NMR) is one of the most versatile techniques with applications beyond the natural sciences. High magnetic field is beneficial for improving sensitivity and enhancing resolution in NMR, because nuclear magnetization and chemical shift are proportional to the strength of an applied magnetic field. The currently highest available static continuous magnetic field is generated using a hybrid magnet and the maximum field achieved does not exceed 45 T [1]. However, in many cases, the strength of this available highest magnetic field is not enough. For example, copper-oxide high-temperature superconducting materials have upper critical fields far higher than those generated by any DC magnets. Therefore, in order to achieve the transition from the low-temperature normal state to the high temperature superconducting state, a higher magnetic field is required. So far, the available highest magnetic field is produced by a pulsed magnet temporarily where the field is time dependent. Pulsed magnets are widely used around the world today, and a maximum pulsed field of more than 100 T was reported in the last few years [2]. For this reason, there is a growing demand for pursuing NMR in pulsed fields. Haase et al. [3], Zheng et al. [4], and Stork et al. [5] have pioneered NMR in pulsed magnetic fields, and observed NMR signals using the free induction decay (FID) and the spin-echo methods in these time dependent fields. All of these experiments employed the conventional pulsed NMR technique, in which a radio frequency (RF) pulse or a pulse sequence was applied to excite nuclei in the vicinity of the maximum of a pulsed field. Therefore, NMR signals could only be detected in some discrete moments near the peak of pulsed fields. Thus, it could not take full advantage of pulsed high magnetic fields, for instance, it is unfeasible to study the whole process of some important reactions, such as field-induced phase transition in materials, chemical synthesis, structural change in biological molecules, etc.

A novel experimental scheme, wideband continuous wave NMR (WB-CW-NMR), which is promising to enable NMR techniques to expand dramatically in the time dimension, is presented in this article. The WB-CW-NMR scheme in pulsed magnetic fields, as illustrated in Fig. 1, is a combination of the traditional field-sweep and frequency-sweep continuous wave NMR methods. It is expected that magnetic resonance will continuously occur when the Larmor frequency $\omega_0$ ($=\gamma B_z$, where $\gamma$ and $B_z$ are the nuclear gyromagnetic ratio and the strength of the applied pulsed magnetic field, respectively) is within the bandwidth of the wideband continuous RF excitation.

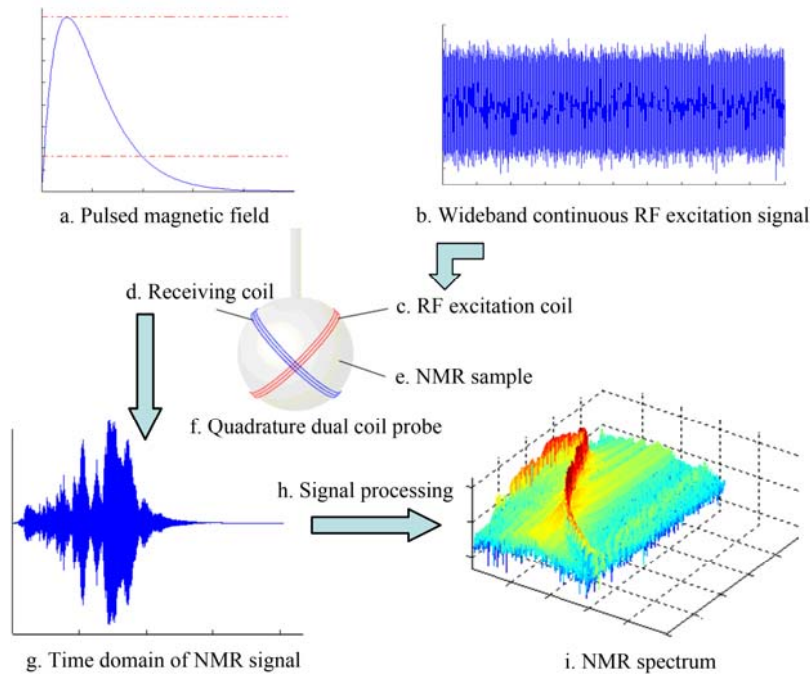

**Fig. 1.** The WB-CW-NMR experimental scheme. A wideband continuous RF signal (**b**) is applied to the RF excitation coil (**c**) in the $x$ direction to excite nuclei in a pulsed magnetic field (**a**) whose direction is always defined as the $z$-axis, and WB-CW-NMR signal (**g**) is detected synchronously by the receiving coil (**d**) in the $y$ direction of the quadrature dual coil probe (**f**). The WB-CW-NMR spectrum (**i**) can be obtained with appropriate signal processing (**h**).

Soon after the experimental discovery of NMR in 1946, a set of phenomenological equations, the so-called Bloch equations, were proposed by Bloch. The solution of the Bloch equations can provide clear insight about the evolution of various magnetization components. Up to now, these equations have been solved by application of a Laplace transform [6], a multiple-derivative method [7], numerical integration [8], Lagrange interpolation [9], etc. Despite the availability of these solutions, as far as we know, there are no solutions under the condition of pulsed magnetic fields, especially when the RF excitation field is a wideband continuous wave.

The purpose of this study is to validate the WB-CW-NMR experimental scheme by solving the modified time-dependent Bloch equations. In addition, the influences of various parameters, such as the digital sequence of wideband continuous RF excitation signal, the strength of applied magnetic field and RF power, on the magnetization $M_y$ are analyzed briefly. Finally, we discuss the main difficulties in verifying the WB-CW-NMR scheme experimentally, and propose a preliminary plan and basic requirements for detecting the WB-CW-NMR signal.

**2**. **Theoretical Consideration**

The classic form of the Bloch equations in the vector notation in the laboratory frame is given by [10]

$$\frac{d\mathbf{M}}{dt} = \gamma \mathbf{M} \times \mathbf{B} - \frac{M_x \mathbf{i} + M_y \mathbf{j}}{T_2} - \frac{M_z - M_{z0}}{T_1} \mathbf{k}, \qquad (1)$$

where $\gamma$ is the nuclear gyromagnetic ratio, $\mathbf{M} = (M_x, M_y, M_z)$ is the total magnetization vector, $T_1$ and $T_2$ are, respectively, the spin-lattice and spin-spin relaxation time, $\mathbf{B}$ describes the magnetic field, which including a static applied magnetic field $B_0$ along the $z$ axis and an RF excitation field $B_1$ along the $x$ axis, $M_{z0}$ denotes $z$ component of the magnetizations on the thermal equilibrium in the absence of RF irradiation, and the value of $M_{z0}$ in high temperature is given by the well-known Curie's law

$$M_{z0} = \frac{N\gamma^2 \hbar^2 I(I+1) B_0}{3kT} = \chi_0 B_0, \qquad (2)$$

where $N$ is the number of nuclei per unit volume, $\hbar$ is the reduced Planck constant, $I$ is the nuclear spin quantum number, $T$ is the absolute temperature, and $\chi_0$ is the static nuclear susceptibility.

However, the physical mechanism of WB-CW-NMR in pulsed fields is more complex than that of continuous wave or pulsed NMR. Firstly, the strength of a pulsed magnetic field changes over time, so the relaxation time and macroscopic magnetization $M_{z0}$ are no longer constants. Moreover, in the initial and end stages of a pulsed field, the RF excitation field is stronger than the applied pulsed magnetic field, thus it is necessary to modify the classical Bloch equations. In the following part, these issues will be discussed in more detail.

*2.1 Relaxation time in pulsed magnetic fields*

The most important aspect of the WB-CW-NMR experimental scheme on theoretical grounds is that enough nuclei should be excited to produce a detectable NMR signal, while avoiding sample saturation. To polarize the nuclei rapidly and prevent sample saturation in pulsed magnetic fields, a moderate amount of paramagnetic irons are routinely added to reduce the relaxation time of liquid samples. Gadolinium is widely used for this purpose because of the favorable magnetic (electron spin relaxation) and coordination (high coordination number and easy access of water to the inner coordination sphere) properties of complexes formed by this ion with a variety of ligands.

The addition of paramagnetic ions to water, as was first shown by Bloch, Hansen, and Packard [11], can influence markedly the proton relaxation time. Soon after, Bloembergen, Purcell, and Pound proposed the so-called Bloembergen-Purcell-Pound (BPP) theory to explain the nuclear spin relaxation in their excellent work [12]. Theoretical and experimental results in their article showed that paramagnetic complexes can increase the relaxation rate ($1/T_1$ and $1/T_2$) of bulk water. Subsequently, this theory was generalized by Solomon in 1955 [13] and extended by Bloembergen and Morgan to include electron spin relaxation in 1961 [14]. Besides being affected by paramagnetic salt, the relaxation time is also related to the magnetic field strength. The frequency dependencies of relaxation time in aqueous solutions of certain type and concentration of paramagnetic ions are given by the Solomon-Bloembergen-Morgan equations [14] for $\omega_I \ll \omega_S$,

$$\frac{1}{T_1} = K_1 \left[ \frac{3\tau_c}{1+\omega_I^2\tau_c^2} + \frac{7\tau_c}{1+\omega_S^2\tau_c^2} \right] + K_2 \left[ \frac{\tau_e}{1+\omega_S^2\tau_e^2} \right], \tag{3}$$

and,

$$\frac{1}{T_2} = \frac{K_1}{2} \left[ 4\tau_c + \frac{3\tau_c}{1+\omega_I^2\tau_c^2} + \frac{13\tau_c}{1+\omega_S^2\tau_c^2} \right] + \frac{K_2}{2} \left[ \tau_e + \frac{\tau_e}{1+\omega_S^2\tau_e^2} \right], \tag{4}$$

In these equations, $\omega_I$ is the nuclear precession frequency, $\omega_S$ is the electron precession frequency, and they are both proportional to the applied magnetic field strength. The condition $\omega_I \ll \omega_S$ always holds because the gyromagnetic ratio of the electron $\gamma_S$ is over 600 times larger than the gyromagnetic ratio of the hydrogen proton $\gamma_I$. $K_1$ and $K_2$ are constants relative to the nuclear gyromagnetic ratio $\gamma_I$, the total electron spin of the metal ion $S$, the nuclear-ion distance $r$, the probability that a proton occupies a position in the hydration sphere of a paramagnetic ion in a 1 mol/L solution of the ions $p$, etc. $\tau_c$ and $\tau_e$ are the correlation times for dipolar and spin exchange interactions, respectively. The value of $\tau_c$ is determined by the rotation of the hydrated ion and is about $10^{-11}$ s [15], and $\tau_e$ is estimated to be on the order of $10^{-12} - 10^{-9}$ s for different ions [16]. In very low external fields ($\omega_S \tau_e \ll 1$), $T_1$ and $T_2$ for protons in paramagnetic solutions are approximately equal. With the increasing of magnetic field strength, the behavior of $T_1$ and $T_2$ appears to fall into several categories for paramagnetic solutions containing different type irons [15]. In addition to paramagnetic ions and magnetic field strength, there are other factors, such as temperature, viscosity, pH, etc., which affect the relaxation time, so the change of relaxation time is unfeasible to predict accurately with a theoretical calculation. In this paper, we assume that the field strength dependence of the relaxation time $T_1$ and $T_2$ can be described by Eqs. (3) and (4) for a liquid NMR sample containing a certain type and concentration of paramagnetic ions, respectively.

*2.2 Modification of Bloch equations when the applied pulsed magnetic field is weak*

Because of the discharge characteristics of a pulsed magnet, the instantaneous strength of a pulsed magnetic field in the initial and end stages are both rather weak. It means that the RF excitation field $B_x$ is higher compared with the applied pulsed field $B_z$ in these moments, that is to say, the condition $B_x \ll B_z$ is not satisfied. Therefore, the validity of the Bloch equations (1) becomes questionable under this condition. These topics were discussed systematically in the classic NMR textbook of Abragam [17]. It was suggested that the assumption of a magnetization relaxing towards the equilibrium value $M_{z0} = \chi_0 B_z$ should be replaced by that of a relaxation towards the instantaneous value $\chi_0(B_z(t) + B_x(t))$. Additionally, in pulsed magnetic fields, macroscopic magnetization $M_{z0}(t)$ will change with the field strength. When the pulsed magnetic field $B_z(t)$ varies sufficiently slowly so that the change of precession phase of the magnetization owing to field variation satisfies the following condition [17],

$$\left|\gamma \frac{dB_z(t)}{dt}\right|\tau_c \ll 1, \tag{5}$$

we assume that, $M_{z0}(t)$ can be calculated using Curie's law with the instantaneous magnetic field strength $B_z(t)$ instead of a constant magnetic field $B_0$.

Therefore, when the applied magnetic field is a pulsed field and RF excitation is a wideband continues oscillating field, the Bloch equations ignoring diffusion and radiation damping can be modified by

$$\begin{cases} \dfrac{dM_x}{dt} = \gamma M_y(t)B_z(t) - \dfrac{M_x(t) - \chi_0 B_x(t)}{T_2} \\ \dfrac{dM_y}{dt} = \gamma(M_z(t)B_x(t) - M_x(t)B_z(t)) - \dfrac{M_y(t)}{T_2} \\ \dfrac{dM_z}{dt} = -\gamma M_y(t)B_x(t) - \dfrac{M_z(t) - M_{z0}(t)}{T_1} \end{cases}, \tag{6}$$

where $B_z(t)$ is the pulsed magnetic field and $B_x(t)$ is the wideband continuous RF excitation field, $M_{z0}(t)$ is the transient equilibrium magnetization at time $t$ in pulsed magnetic fields. Equation (5), nevertheless, may be not satisfied at the beginning of the following numerical calculation where the pulsed magnetic field changes with time rapidly. This modified Bloch equations used here to describe system evolution even though they are not quite accurate in the initial stage.

Equation (6) is a set of mutually coupled, inhomogeneous linear differential equations. Because the coefficient terms of these equations are time-dependent, and there is no suitable mathematical model to describe the wideband continuous RF excitation signal, it is difficult to give the analytical solutions directly, but they could be solved easily using a numerical integration method. Variable coefficients mean that characteristic values of the coefficient matrix are no longer constant. Consequently, the fast numerical method for solving Bloch equations using a matrix operation [18] is not suitable. In addition, the digital sequences of the wideband continuous RF excitation must be generated in advance, which means that the minimum time step for numerical integration have been determined. Therefore, some variable-step methods for solving differential equations could not be applied. In this article, the numerical solutions of Eq. (6) will be given using the classical fourth-order Runge-Kutta (RK-4) method.

**3. Simulation parameters and result**

The sample used in our simulation is an $H_2O$ solution containing the appropriate amount of $Gd^{3+}$ paramagnetic irons. We assume the correlation times $\tau_c$ and $\tau_e$ are $2\times10^{-11}$ s and $10^{-9}$ s, respectively. The spin-lattice relaxation time $T_1$ and the spin-spin relaxation time $T_2$ of $^1H$ are 500 μs and 200 μs [19] in high magnetic field (~62 T), respectively, and they are equal when the applied magnetic field $B_z$ is smaller than $10^{-3}$ T. The variation of relaxation time $T_1$ and $T_2$ with field strength used in our simulation is shown in Fig. 2. For the $^1H$ nucleus, the spin quantum number is $I = 1/2$ and gyromagnetic ratio $\gamma$ is $2.675\times10^8$ rad·s$^{-1}$·T$^{-1}$. The sample volume (1 mm diameter, 2 mm length) is about 1.6 μL. The number $N$ of hydrogen atoms per volume in aqueous solution is about $6.69\times10^{28}$ m$^{-3}$. The experiment temperature $T$ is set at 300 K.

The initial values of Eq. (6) at $t = 0$ are determined by the initial conditions of the WB-CW-NMR experiments. Since the NMR sample is not magnetized at the beginning of a pulsed magnetic field, the initial macroscopic magnetizations in the following simulation are $M_{z0}(t) = 0$ and $\mathbf{M}(0) = (0,0,0)$.

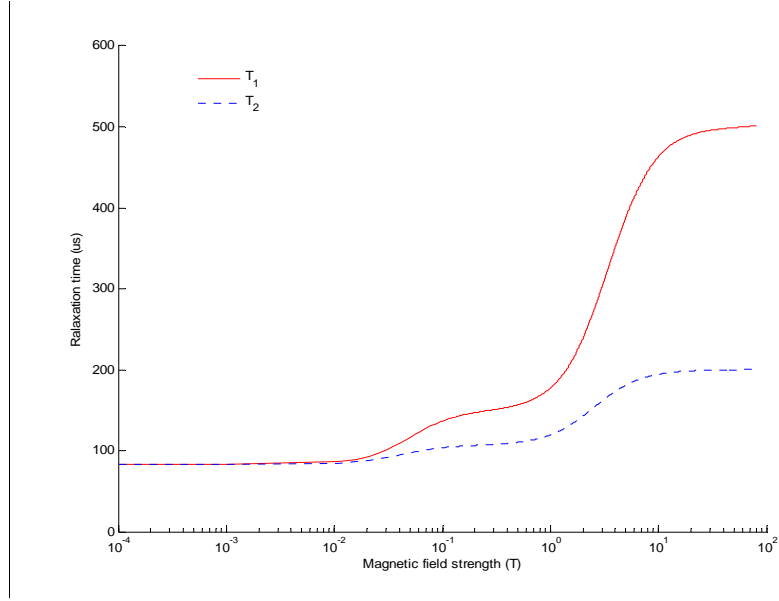

**Fig. 2** The dependence of relaxation time of $^1$H on magnetic field strength in water solutions of specific concentration Gd$^{3+}$ ions. The coefficients $K_1$ and $K_2$ in Eqs. (3) and (4) are $3.33 \times 10^{13}$ and $5.33 \times 10^{12}$, respectively.

The waveform of a pulsed magnetic field (see Fig. 3) can be described using a double exponential model

$$B_z(t) = kB_{zm}(e^{-\alpha t} - e^{-\beta t}), \tag{7}$$

where $B_{zm}$ is the maximum of a pulsed magnetic field, $\alpha$ and $\beta$ are parameters for adjusting the pulse peak time, cutting-edge, and half-width etc. $t_{\max}$ and $k$ are, respectively, the time and correction factor of the maximum magnetic field $B_{zm}$, which are given by

$$t_{\max} = \frac{\log(\alpha) - \log(\beta)}{\alpha - \beta}, \tag{8}$$

and,

$$k = \frac{1}{e^{-\alpha t_{\max}} - e^{-\beta t_{\max}}}. \tag{9}$$

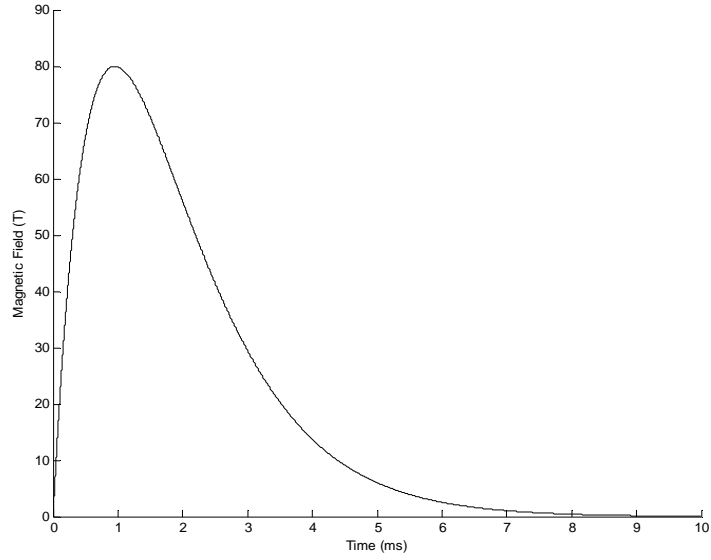

**Fig. 3** The waveform of a pulsed magnetic field. The parameters are $B_{zm}$ = 80 T, $\alpha$ = 1000 s$^{-1}$, and $\beta$ = 1100 s$^{-1}$. Under these conditions, the duration of the pulsed magnetic field is $t_d = 10$ ms and reaches a maximum at $t_{max} = 0.95$ ms

Figure 4 shows the digital sequence of wideband continuous RF excitation, which is generated using white noise through an appropriate digital bandpass filter. The corresponding frequency of the maximum of the pulse magnetic field $B_{zm}$ is set as $f_0 \, (= \gamma B_{zm}/2\pi \approx 3.40$ GHz). The parameters of the bandpass filter are chosen as follows: the pass band is 0.2–1.0 $f_0$ (0.68–3.40 GHz), the ripple within the pass band is 0.01 dB, the transition bands are 0.1–0.2 $f_0$ (0.34–0.68 GHz) and 1.0–1.1 $f_0$ (3.40–3.74 GHz) with an attenuation of 80 dB, and the sampling rate is 100 GHz. The mean-square-root (RMS) of wideband continuous RF excitation field $B_{xr}$ is set to 0.01 T.

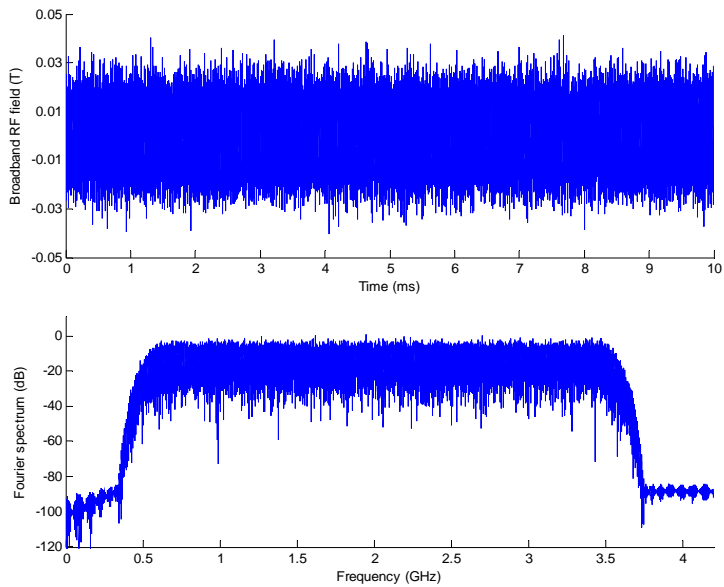

**Fig. 4** The time-domain waveform (upper) and Fourier amplitude spectrum (lower) of the wideband continuous RF excitation

The wideband continuous RF signal is a novel excitation, which is promising to be applied to NMR experiments in pulsed magnetic fields. Unlike the pulsed NMR experimental scheme, in which FIDs are merely detected in some discrete moments, NMR signal is expected to be detected continuously in the WB-CW-NMR experimental scheme. The simulation result of magnetization $M_y$ in the laboratory frame is given in Fig. 5. The waveform of WB-CW-NMR signal is different from FIDs in pulsed NMR experiments. The NMR signal appears from 0.1 ms to 5.8 ms, and reaches a maximum at 3.5 ms. The WB-CW-NMR signal duration is in accordance with the effective time period of the pulsed magnetic field, during which the Larmor frequency is within the bandwidth of the wideband continuous RF excitation.

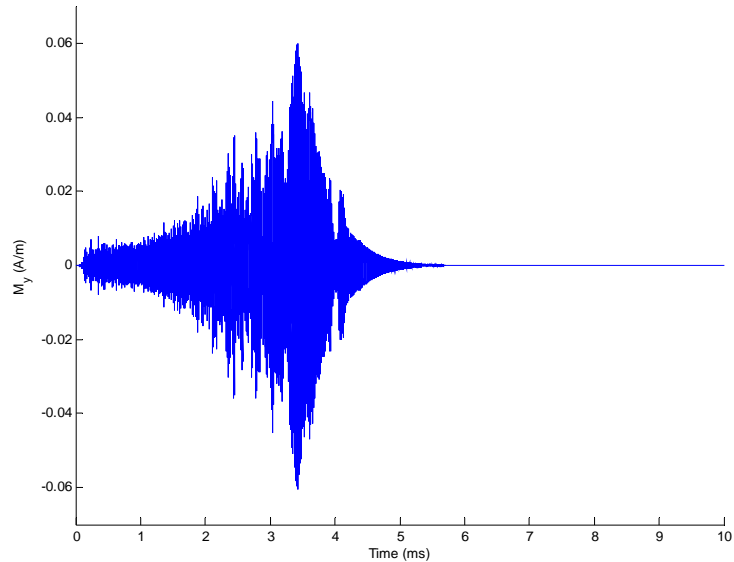

**Fig. 5** Simulated magnetization $M_y$ in the laboratory frame in the case of a wideband continuous RF irradiation in a pulsed magnetic field up to 80 T.

### 3.1 Effect of wideband continuous RF sequences

The white noise sequence used in our numerical simulation is a column of uniformly distributed pseudorandom numbers generated by the function 'Rand' in Matlab. For a certain length random sequence, its value is completely determined by the random seed. In order to investigate the effect of wideband continuous RF sequences on WB-CW-NMR signal, we use different random seeds to generate four pseudorandom sequences, and give the waveform of magnetization $M_y$ in Fig. 6. The RMS values of these RF sequences are equal. The simulated results show that, the waveform details of these NMR signals are different, but the amplitude and duration are roughly the same. In consequence, the choice of a random sequence is not a key factor for the WB-CW-NMR experiment. In order to eliminate the effects of different random sequences on simulation results, the wideband continuous RF excitation signal is generated by filtering the identical random number sequence in the following simulation.

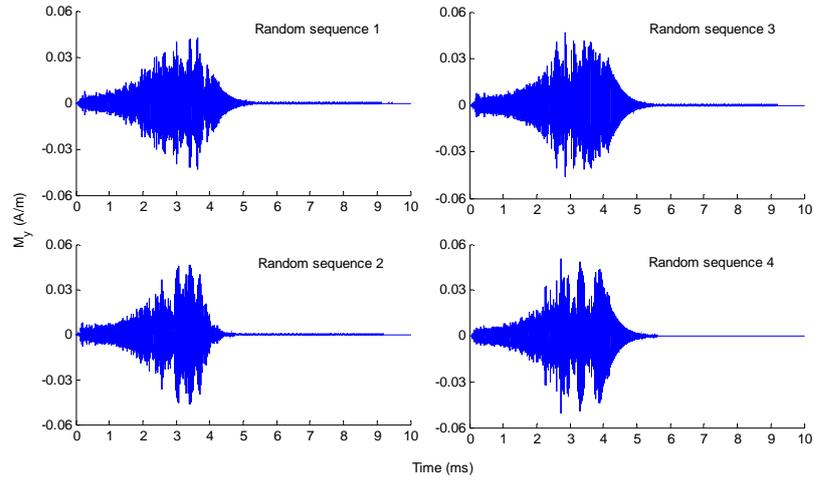

**Fig. 6** The magnetization $M_y$ in the laboratory frame as a function of time for different wideband continuous RF sequences in a pulsed magnetic field up to 80 T.

## 3.2 Effect of applied magnetic field strength

The transient equilibrium magnetization $M_{z0}(t)$ is determined by the strength of the applied magnetic field, and the relaxation time of the atomic nuclei is also closely related to the magnetic field strength. Figure 7 shows the influence of pulse magnetic field strength on the WB-CW-NMR signal. The duration and peak time of these pulsed magnetic fields are all the same ($t_d = 10$ ms, and $t_{max} = 0.95$ ms), whereas the peak value $B_{zm}$ increases from 10 T to 80 T with a step of 10 T. The RMS values of all wideband continuous RF excitations are 0.01 T.

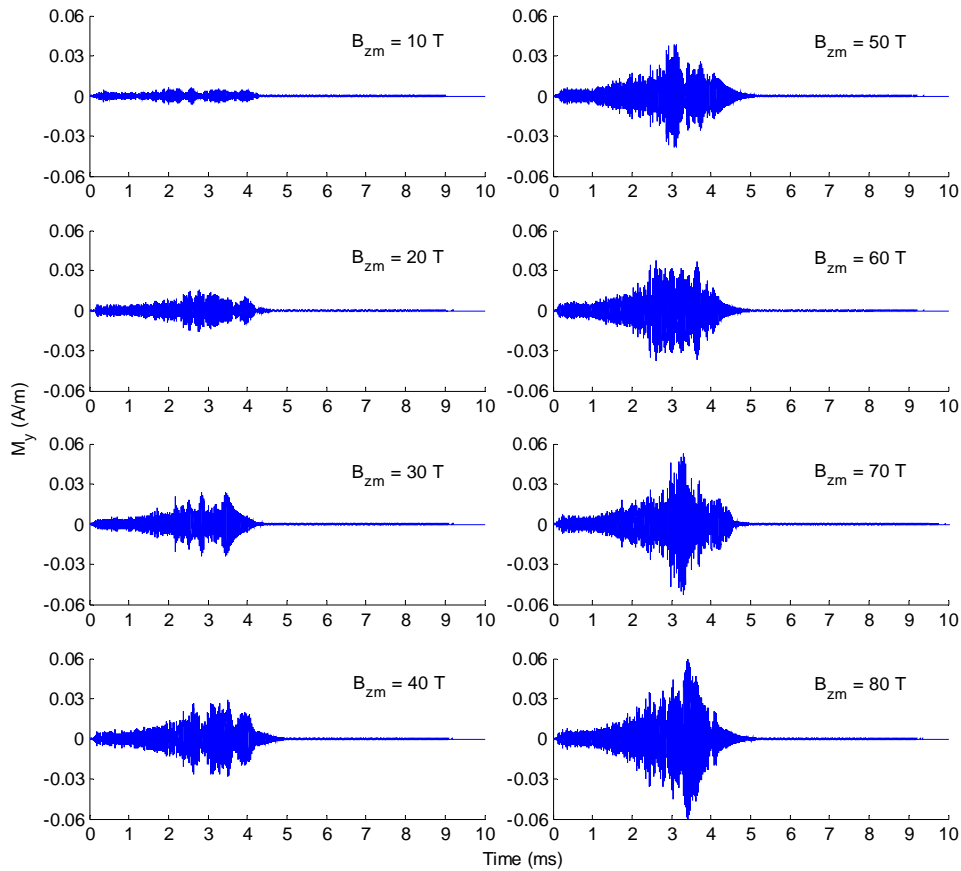

**Fig. 7** The magnetization $M_y$ in the laboratory frame as a function of time for different peak values of pulsed magnetic field in the case of identical wideband continuous RF irradiation. The maximum of pulsed magnetic field $B_{zm}$ are 10, 20, 30, 40, 50, 60, 70, 80 T, respectively.

Due to polarization intensity and relaxation time, there are obvious differences between these WB-CW-NMR signal waveforms in different magnetic field strength. The amplitude of the NMR signal increases gradually with the increasing of the strength of the pulse magnetic field, which is approximately linear (Fig. 8). This suggests that, in the WB-CW-NMR scheme, high magnetic field will also be beneficial for enhancing the NMR signal.

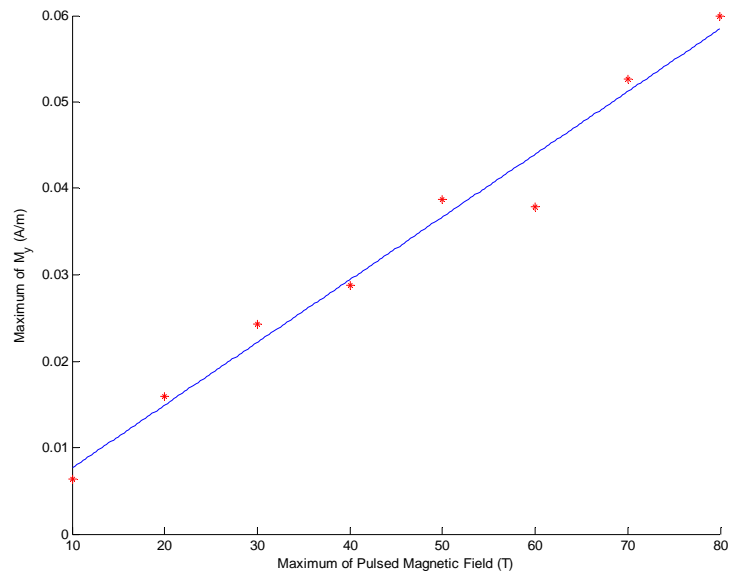

**Fig. 8** Dependence of the maximum of magnetization $M_y$ on the maximum of pulsed magnetic field $B_{zm}$. Red asterisks are simulation results, and the blue line is the fitting curve using a linear fitting method.

*3.3 Effect of wideband continuous RF power*

The magnetization $M_y$ trajectories for different RF power are shown in Fig. 9. The figure indicates that the RF excitation field strength directly determines the amplitude of the NMR signal. The strength of the NMR signal initially increases with RF power and reaches a maximum when RF excitation power is sufficiently high, but subsequently decreases at higher RF power, suggesting that the sample has been saturated. Consequently, selecting an appropriate RF power is a decisive factor in the WB-CW-NMR experiment.

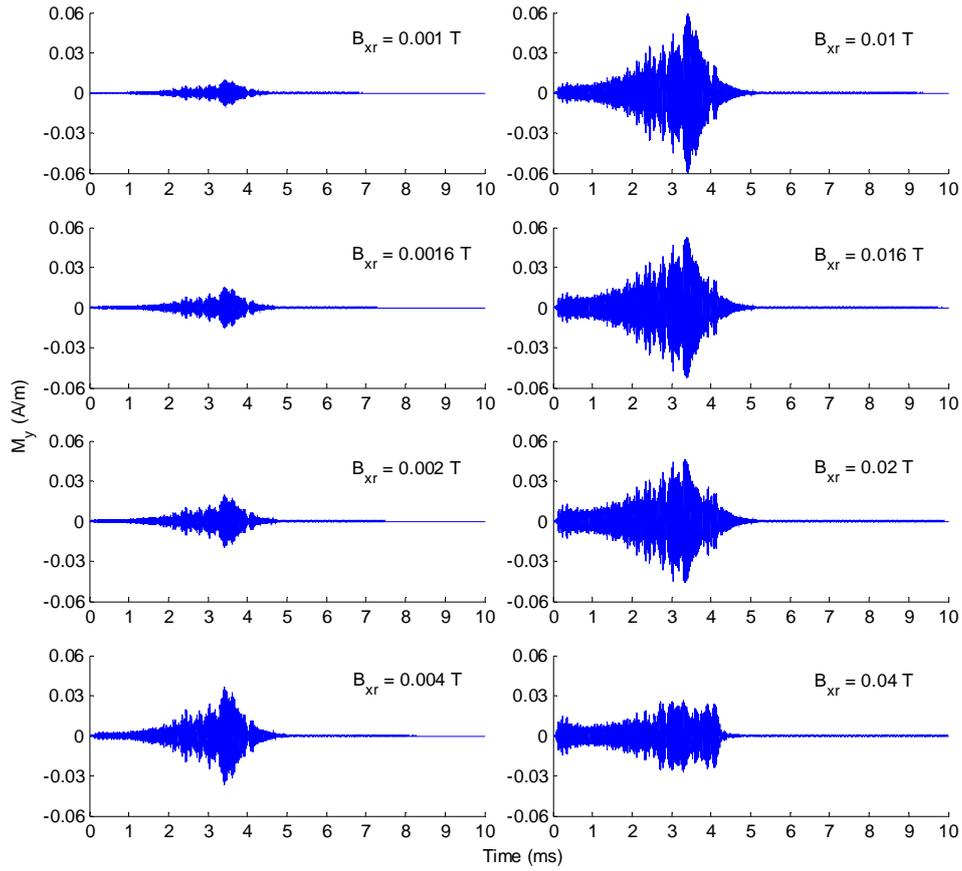

**Fig. 9** The magnetization $M_y$ in the laboratory frame as a function of time for different RF excitation strength in a pulsed magnetic field up to 80 T. The RMS of wideband continuous RF excitation signal $B_{xr}$ are 0.001, 0.0016, 0.002, 0.004, 0.01, 0.016, 0.02, 0.04 T, respectively.

Since the bandwidth of the excitation signal is rather wide, the required RF power in WB-CW-NMR experimental scheme is correspondingly larger than that in traditional CW-NMR. Fig. 10 shows that the optimal RF strength is approximately 0.01 T (RMS) in our simulation condition.

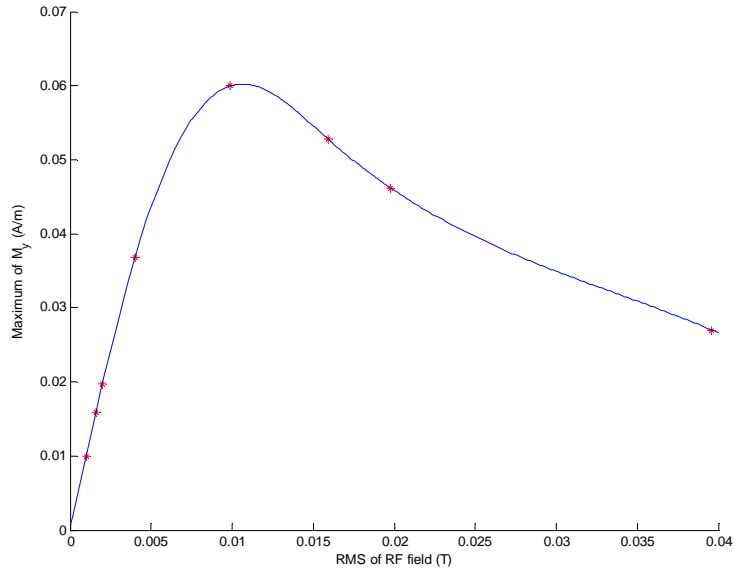

**Fig. 10** Dependence of the maximum of magnetization $M_y$ on RF excitation strength $B_{xr}$. Red asterisks are simulation results, and the blue line is the fitting curve using a spline-interpolation method.

## 4. Discussion

In this section, we will discuss the feasibility of detecting the WB-CW-NMR signal experimentally. The problem which should be solved primarily is how to meet the ultra-wideband requirement of the WB-CW-NMR experimental system. Recently, the bandwidth of tuneless wideband NMR probes constructed by transmission lines has been able to reach tens and even hundreds of MHz [20]. Further expansion of the probe bandwidth can employ the multi-resonance probe technique [21]. The wideband continuous RF excitation signal can be produced using the technique of direct digital synthesis (DDS). The bandwidth of wideband RF signal is mainly limited by the performance of the digital-to-analog converter (DAC). To take both high quantization precision and large dynamic range into account, it is very difficult to produce an ultra-wideband signal, say 3 GHz, in the hardware framework of a single transmitting/receiving channel. We intend to employ the intersection of frequency spectrum technique to achieve ultra-wideband RF signal transmission and reception. The Multi-channel WB-CW-NMR experiment system is shown in Fig. 11. Each channel in this Multi-channel system only processes a signal within a particular, relatively narrow frequency band. A DDS signal generator produces a certain bandwidth RF continuous signal. After being smoothed with a low-pass filter (LPF), this baseband signal is modulated with different RF frequencies, and then passed through a band-pass filter (BPF) and a power amplifier (PA). The required ultra-wideband signal is synthesized using these wideband signals in different frequency bands. Similarly, the WB-CW-NMR signal is divided into multiple channels by a demultiplexer. After low noise amplification, demodulation, low pass filtering and ADC, the data of each channel is processed individually, and then integrated to obtain the spectrum for the entire ultra-wideband frequency band.

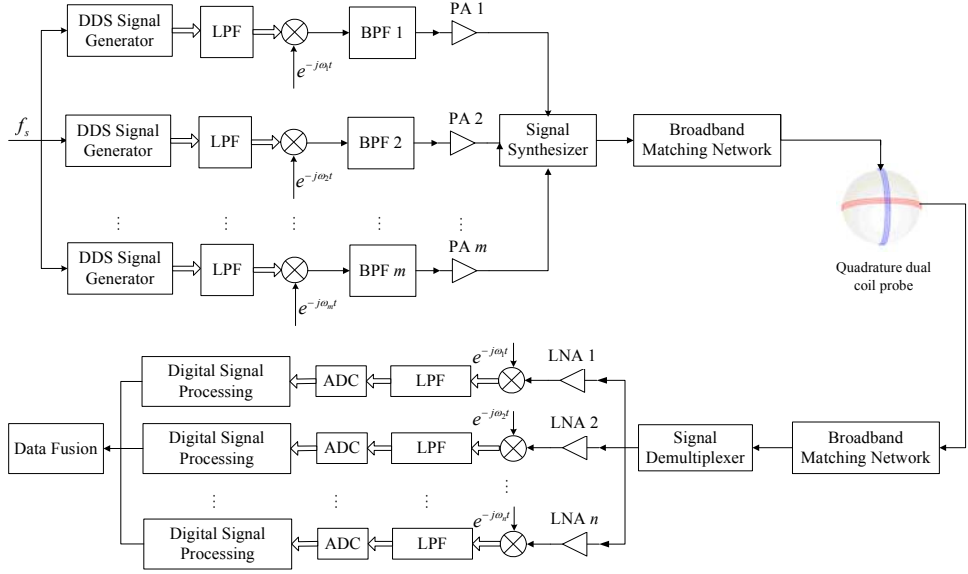

**Fig. 11** Schematic diagram of WB-CW-NMR experimental system based on intersection of frequency spectrum.

However, for a multi-channel NMR system, there are many new issues which need further consideration, such as how to ensure that the system has good amplitude-frequency, phase-frequency and nonlinear distortion characteristics over the entire ultra-wideband frequency range, and how to improve the consistency, while reducing mutual interference between different channels. Of course, there are some other compromise solutions, such as decreasing magnetic field strength or selecting nuclei with a low gyromagnetic ratio, to reduce bandwidth requirements.

It is actually not magnetization $M_y$, but the rate of change of the rapidly varied nuclear induction $B_y = \mu_0 M_y$, which is observed in the experiment. The induced electromotive force (IEF) $\varepsilon$ received by the receiving coil in the *y* direction is

$$\varepsilon = -N_c S \frac{dB_y}{dt} = -\mu_0 N_c S \frac{dM_y}{dt}, \tag{10}$$

where $\mu_0$ is the magnetic permeability of free space, $N_c$ is the number of turns on the receiver coil, and *S* is the effective cross-sectional area of the receiving coil. Assuming that the number of turns $N_c$ is 10, the effective area of the coil *S* is $0.785\,\mathrm{mm}^2$ (1 mm diameter). The IEF received by the receiving coil, which is calculated according to Eq. (10), is shown in Fig. 12. The term $dM_y/dt$ is obtained by the difference of magnetization $M_y$, i.e. $\Delta M_y/\Delta t$. The waveform of $M_y$ used here is shown in Fig. 5. The maximum voltage induced in the receiving coil is approximately 4 mV.

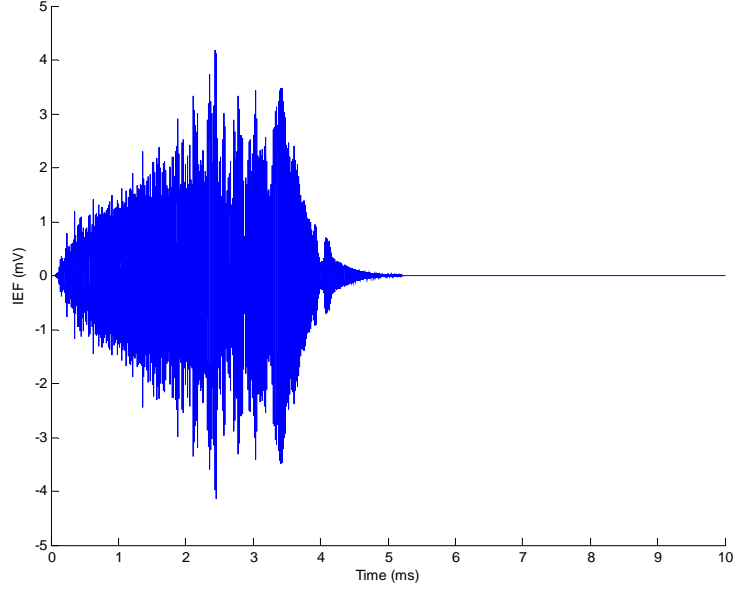

**Fig. 12** The IEF received by the receiving coil under a wideband continuous RF irradiation in a pulsed magnetic field up to 80 T. The values of some parameters used in estimating the amplitude of the IEF are listed as follows: $I = 1/2$, $\gamma = 2.6752 \times 10^8$ rad·s$^{-1}$·T$^{-1}$, $N = 6.6889 \times 10^{28}$ m$^{-3}$, $B_{zm} = 80$ T, $B_{xr} = 0.01$ T, $T = 300$ K.

One application of the WB-CW-NMR scheme is to achieve the measurement of pulsed magnetic field over a wide range. Figure 13 shows the time-frequency spectrum of the IEF using a short-time Fourier transform. The strength of the magnetic field $B_z(t)$, which can be calculated according to the Larmor relationship ($B_z(t) = 2\pi f(t)/\gamma$, where $f(t)$ is the frequency of WB-CW-NMR signal at time $t$), is in agreement with the waveform of the pulsed magnetic field (as shown in Fig. 3), indicating the validity of WB-CW-NMR experimental scheme.

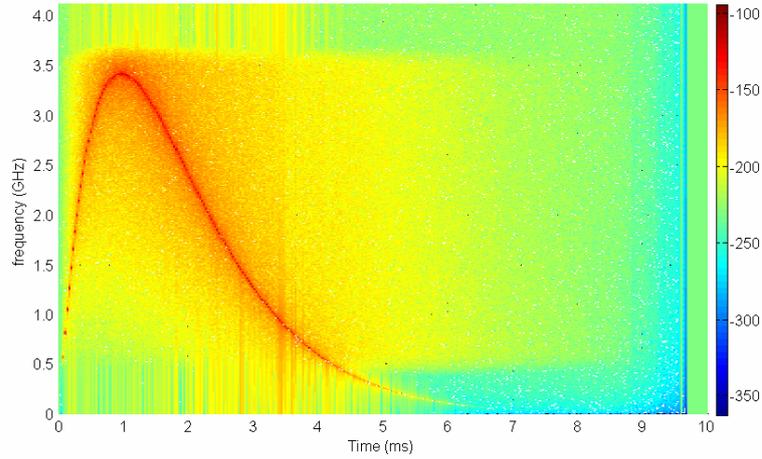

**Fig. 13** The time-frequency spectrum of the IEF using a short time Fourier transform.

In an ideal system, the measured noise is set only by the thermal Johnson noise, explained by Nyquist. The noise is generated by the resistance of the probe coil. The relationship between noise voltage $V_n$ and experimental temperature $T$, bandwidth $\Delta f$, and resistance of the coil $R$ is given by

$$V_n = \sqrt{4k_B T \Delta f R}, \tag{11}$$

where $k_B$ is the Boltzmann constant. Assuming an experimental temperature $T$ = 300 K, the bandwidth $\Delta f$ = 150 MHz for each channel, and the resistance $R$ is 10 Ω, the calculated value of noise voltage $V_n$ is about 4.98 μV. It is far below the amplitude of IEF, indicating that there is potentially a high signal-to-noise ratio (SNR) in the WB-CW-NMR experiment.

In the WB-CW-NMR experiment, the transmission of a wideband continuous RF excitation signal and reception of WB-CW-NMR signal are performed synchronously. The main problem in the WB-CW-NMR experiment is the detection of a weak signal (maximum of $6 \times 10^{-2}$ A/m) in the presence of a larger wideband continuous RF magnetic field (about $2.4 \times 10^4$ A/m). Because of orientation errors of the quadrature coil, the actual isolation of dual coils is often limited. Precision machining and orientation adjustment could make the leakage of the RF excitation field into the receiver coil reduce by a factor of $10^4$ or thereabouts [11]. However, this leakage, which is also called RF interference, is still one to two orders higher than the NMR signal on detection. Therefore, it is not easy to distinguish WB-CW-NMR signal from this large background, which is another main difficulty in verifying the WB-CW-NMR scheme experimentally. Nevertheless, RF interference and noise could be cancelled by some adaptive algorithms [22]. It is entirely possible to extract the weak WB-CW-NMR signal from raw data if the effect of the cancellation algorithm could reach more than

30 dB.

The basic set-up for an adaptive interference canceling system is shown in Fig. 14. The primary signal is the raw NMR data, $y(k)$, which is comprised of WB-CW-NMR signal $s(k)$, RF interference $i_1(k)$ and noise $n_1(k)$; the reference signal $x(k)$ is generated by the identical digital sequence passing through the same NMR system without applying a pulse magnetic field, which means there are only RF interference $i_2(k)$ and noise $n_2(k)$, but no NMR signal. If these two signals are correlated, the adaptive algorithm can adjust the filter to make the output $d(k)$ as similar to $i_1(k)+n_1(k)$ as possible by minimizing the error signal, $e(k)$. In order to effectively suppress the large RF interference, the correlation between these two interferences $i_1(k)$ and $i_2(k)$ should be strong.

The linear correlation between two signals $y(k)$ and $x(k)$ is measured as a function of frequency by the magnitude squared coherence function

$$\left|\gamma_{xy}(f)\right|^2 = \frac{\left|P_{xy}(f)\right|^2}{P_{xx}(f)P_{yy}(f)} \quad (12)$$

Here, $P_{xy}(f)$ is the complex cross-power spectral density, $P_{xx}(f)$ and $P_{yy}(f)$ are the power spectral densities of the individual signals. The maximum possible attenuation of an adaptive noise canceller as a function of frequency is given by [23]

$$\text{Attenuation}(f) = -10\log_{10}(1-\left|\gamma_{xy}(f)\right|^2), \quad (13)$$

which means, an attenuation of 30 dB can only be obtained if the correlation between interferences $i_1(k)$ and $i_2(k)$ is larger than 0.999. Therefore, it requires that the WB-CW-NMR system status should be consistent between these two data acquisition; in other words, the stability of the NMR experimental system is another important requirement.

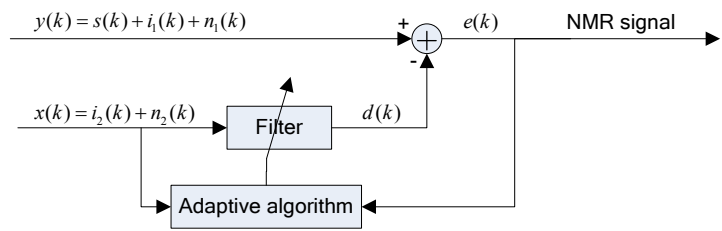

**Fig. 14** Schematic overview of adaptive interference cancellation in WB-CW-NMR data processing.

Furthermore, the SNR of the WB-CW-NMR experiment will degrade because of poor uniformity and rapid changes, which produce eddy effects, of pulsed magnetic fields. The effect of radiation damping [24], which further decreases the SNR, also should be taken into account in high magnetic field. In summary, it is undoubtedly a great challenge to observe a WB-CW-NMR signal experimentally.

## 5. Conclusion

In this paper, we proposed a novel NMR scheme, WB-CW-NMR, and solved the modified Bloch equations using a numerical method, where the applied magnetic field is a pulsed magnetic field up to 80 T and the wideband continuous radio frequency (RF) excitation is a band-limited (0.68–3.40 GHz) white noise. The simulation results verified the effectiveness of our proposed WB-CW-NMR scheme, and showed that the WB-CW-NMR signal is approximately linearly enhanced with the increase of the pulsed magnetic field strength. The power of RF excitation has an optimal value for exciting nuclei effectively and avoiding saturation, and there seems to be no significant relationship between WB-CW-NMR signal strength and wideband continuous RF sequences.

In order to transmit and collect an ultra-wideband signal with low distortion and wide dynamic range, we presented a multi-channel WB-CW-NMR system framework and gave some basic system requirements. We also estimated the magnitude of WB-CW-NMR signal, thermal noise, and RF interference. The results indicated that, the main difficulty in verifying the WB-CW-NMR scheme experimentally comes from the inherent low signal-to-interference ratio (SIR). In order to suppress the large RF interference, we gave a preliminary adaptive interference cancellation plan.

In conclusion, in order to detect the WB-CW-NMR signal experimentally, the isolation of the orthogonal dual coil probe and the effect of the adaptive algorithm should be improved such that the leakage of the wideband continuous RF excitation is reduced by more than five orders of magnitude.

**Acknowledgments** This work was supported by the National Natural Science Foundation of China (Grant No. 11475067), the Innovative Research Foundation of Huazhong University of Science and Technology (Grant No. 2015ZDTD017), and the Experimental Apparatus Research Project of Wuhan Pulsed High Magnetic Field Center (Grant No. 2015KF17).